\documentclass[letter]{aa} 
\usepackage[]{natbib}
\usepackage{graphicx,color}
\usepackage{txfonts}
\usepackage[colorlinks=true, citecolor=blue]{hyperref}

\begin{document} 

\title{Aluminum Enhanced Metal-Poor Stars buried in the Inner Galaxy}

	\author{
		Jos\'e G. Fern\'andez-Trincado\inst{1, 2}\thanks{Corresponding author: jose.fernandez@uda.cl},
		Timothy C. Beers\inst{3},
		Dante Minniti\inst{4,5},
		Baitian Tang\inst{6},
		Sandro Villanova\inst{7},
		Doug Geisler\inst{7,8,9}, 
		Angeles P\'erez-Villegas\inst{10}
		\and
		Katherine Vieira\inst{1}
		}
	
	\authorrunning{J. G. Fern\'andez-Trincado et al.} 
	
\institute{
		Instituto de Astronom\'ia y Ciencias Planetarias, Universidad de Atacama, Copayapu 485, Copiap\'o, Chile
		\and
		Institut Utinam, CNRS-UMR 6213, Universit\'e Bourgogne-Franche-Compt\'e, OSU THETA Franche-Compt\'e, Observatoire de Besan\c{c}on, BP 1615, 251010 Besan\c{c}on Cedex, France
		\and
	    Department of Physics and JINA Center for the Evolution of the Elements, University of Notre Dame, Notre Dame, IN 46556, USA
        \and	
        Depto. de Cs. F\'isicas, Facultad de Ciencias Exactas, Universidad Andr\'es Bello, Av. Fern\'andez Concha 700, Las Condes, Santiago, Chile
        \and
        Vatican Observatory, V00120 Vatican City State, Italy
        \and
        School of Physics and Astronomy, Sun Yat-sen University, Zhuhai 519082, China
        \and
        Departamento de Astronom\'\i a, Casilla 160-C, Universidad de Concepci\'on, Concepci\'on, Chile
        \and
        Departamento de Astronom\'ia, Universidad de La Serena, Avenida Juan Cisternas 1200, La Serena, Chile
        \and
        Instituto de Investigaci\'on Multidisciplinario en Ciencia y Tecnolog\'ia, Universidad de La Serena. Benavente 980, La Serena, Chile
        \and
        Universidade de S\~ao Paulo, IAG, Rua do Mat\~ao 1226, Cidade Universit\'aria, S\~ao Paulo 05508-900, Brazil     
    }
	
	\date{Received ...; Accepted ...}
	\titlerunning{Fossil relics of early Milky Way}
	
	
	\abstract
	{Stars with higher aluminum and nitrogen enrichment are often the key pieces for the chemical makeup of multiple populations in almost all globular clusters (GCs). There is also compelling observational evidence that some Galactic components could be partially built from dissipated GCs. Thus, the identification of such kinds of stars among metal-poor field stars may provide insights on the composite nature of the Milky Way (MW) bulge and inner stellar halo, as well as reveal other chemical peculiarities. Here, based on APOGEE spectra, we report the discovery of 29 mildly metal-poor ([Fe/H]$\lesssim-0.7$) stars with stellar atmospheres strongly enriched in aluminum (Al-rich stars: [Al/Fe]$\gtrsim+0.5$), well above the typical Galactic levels, located within the Solar radius toward the bulge region, which lies in highly eccentric orbits ($e\gtrsim0.6$). We find many similarities for almost all of the chemical species measured in this work with the chemical patterns of GCs, so we conjecture that they have likely been dynamically ejected into the bulge and inner halo from GCs formed in situ and/or GC formed in different progenitors of known merger events experienced by the MW, such as the \textit{Gaia}-Sausage-Enceladus and/or Sequoia.
	}
	
	\keywords{stars: abundances -- stars: chemically peculiar -- globular clusters:general -- techniques: spectroscopic}
	\maketitle
	
	\section{Introduction}
	\label{section1}
	
     The current standard picture of the inner region of our Galaxy postulates that it is made up by a complex variety of stellar populations, each with characteristic structure, chemistry, and kinematics \citep[see, e.g.,][and references therein]{Recio-Blanco2017, Queiroz2020b, Rojas-Arriagada2020} belonging to the disc, a massive bar structure ($\sim 10^{10}$ M$_{\odot}$) \citep{Bland-Hawthorn2016}, a \textit{pseudo-bulge}, and a possible \textit{classical} bulge \citep[e.g.,][]{Combes1990, Minniti1996, Athanassoula2005, Zoccali2008, Barbuy2018}, likely to be the product of merger and accretion of primordial GCs or dwarf galaxies \citep[e.g.,]{Belokurov2018, Koppelman2019, Massari2019, Souza2020, Naidu2020}, not to mention the inner halo. While our knowledge about the nature of the bulge has improved substantially in recent years, some key questions remain. In particular, some of the stars seen in the bulge today in the lower-metallicity range ($-2.0<$[Fe/H]$<-0.7$) exhibit dissimilar chemical compositions and kinematics than the more metal-rich stars, and possibly retain information on the earliest phases of formation and evolution of the inner Galaxy \citep[see, e.g.,][]{Fernandez-Trincado2019a, Fernandez-Trincado2019b, Fernandez-Trincado2020}. 
     
     A major advancement in revealing the nature of the low-metallicity stars ([Fe/H]$\lesssim -0.7$) in the bulge region was achieved by the discovery of a large population of giant stars with nitrogen over-abundances \citep[N-rich;][]{Schiavon2017, Fernandez-Trincado2019b} and low-$\alpha$ stars \citep[see, e.g.,][]{Recio-Blanco2017}, which mimic the typical chemical patterns only seen in the so called \textit{second-generation}\footnote{\textit{Second-generation} (2G) is used here to refers to stars in GCs that display altered light-element abundances (e.g., He, C, N, O, Na, Al, and Mg), which are different to those of typical MW field stars.} GC stars. The origin of the N-rich stars (hereafter NRS) remains controversial and still under debate \citep[see][for an alternative view]{Bekki2019}. Recent studies propose the existence of stars that have lower aluminum enrichments ([Al/Fe]$< +0.2$) within $\sim 3.5$ kpc of the Galactic Centre \citep{Barbuy2018}, with chemical patterns that resemble low-mass satellite galaxies of the MW \citep[][]{Das2020}. 
     
     Taking advantage of high-resolution near-IR spectroscopy from the Apache Point Observatory Galactic Evolution Experiment \citep[APOGEE-2;][]{Majewski2017}, we select some key elements (N, Al, Mg, Si, among others) for investigation of any possible chemical peculiarities to help constrain scenarios of their origin and relation to the mass assembly history of the Galaxy \citep[see, e.g.,][]{Fernandez-Trincado2016, Fernandez-Trincado2017, Fernandez-Trincado2019a, Fernandez-Trincado2019b, Fernandez-Trincado2019c, Fernandez-Trincado2019d}. 
     
     In this Letter, we report the discovery of a unique collection of aluminum-enhanced (Al-rich) stars among metal-poor ones, exhibiting Al abundances well-above typical Galactic levels ([Al/Fe]$\gtrsim+0.5$) over a range of metallicities, resembling those long known to exist in Galactic GCs \citep[see, e.g.,][]{Szabolcs2020}, but unlikely to be linked to the accretion of dwarf galaxies, for which larger Al abundance ratios have not been observed to date \citep[][and references therein]{Hasselquist2019}. Observations and sample selection is described in Section \ref{section2}. Results are discussed in Section \ref{section3}.  Finally, our conclusions are summarized in Section \ref{section4}.
	
	\section{Observations}
	\label{section2}
	
    This work makes use of data primarily from the APOGEE-2, which have collected high-resolution ($R\sim$22,500) \textit{H}-band spectra (near-IR, $\sim$15145 $\AA{}$ to 16960 $\AA{}$, vacuum wavelengths) for almost 470,000 sources in their sixteenth data release \citep[APOGEE DR16:][]{Ahumada2020}, as part of the Sloan Digital Sky Survey IV \citep[][]{Blanton2017}. For details on the sample selection, see Appendix \ref{sampleobservations}. 
    
	Since we are focused on selecting dissolved and/or evaporated GC stars in [Al/Fe] space, we selected stars among metal-poor ones with aluminum enrichments larger than that expected in dwarf galaxies \citep[see, e.g.,][]{Hasselquist2019} and MW stars, i.e., [Al/Fe]$\gtrsim+0.5$. Our final sample of Al-rich stars comprises 29 stars. With the uncalibrated \texttt{ASPCAP} stellar parameters fixed (T$_{\rm eff}$, $\log$ \textit{g}, and [M/H]), we derive (when possible) the abundances of C, N, O, Mg, Al, Si, and Ce, as well as the metallicity for each Al-rich star. The derived abundances and kinematic parameters for this sample are presented in Tables \ref{Table1} and \ref{Table2}.
	
\begin{figure*}	
	\begin{center}
		\includegraphics[width=190mm]{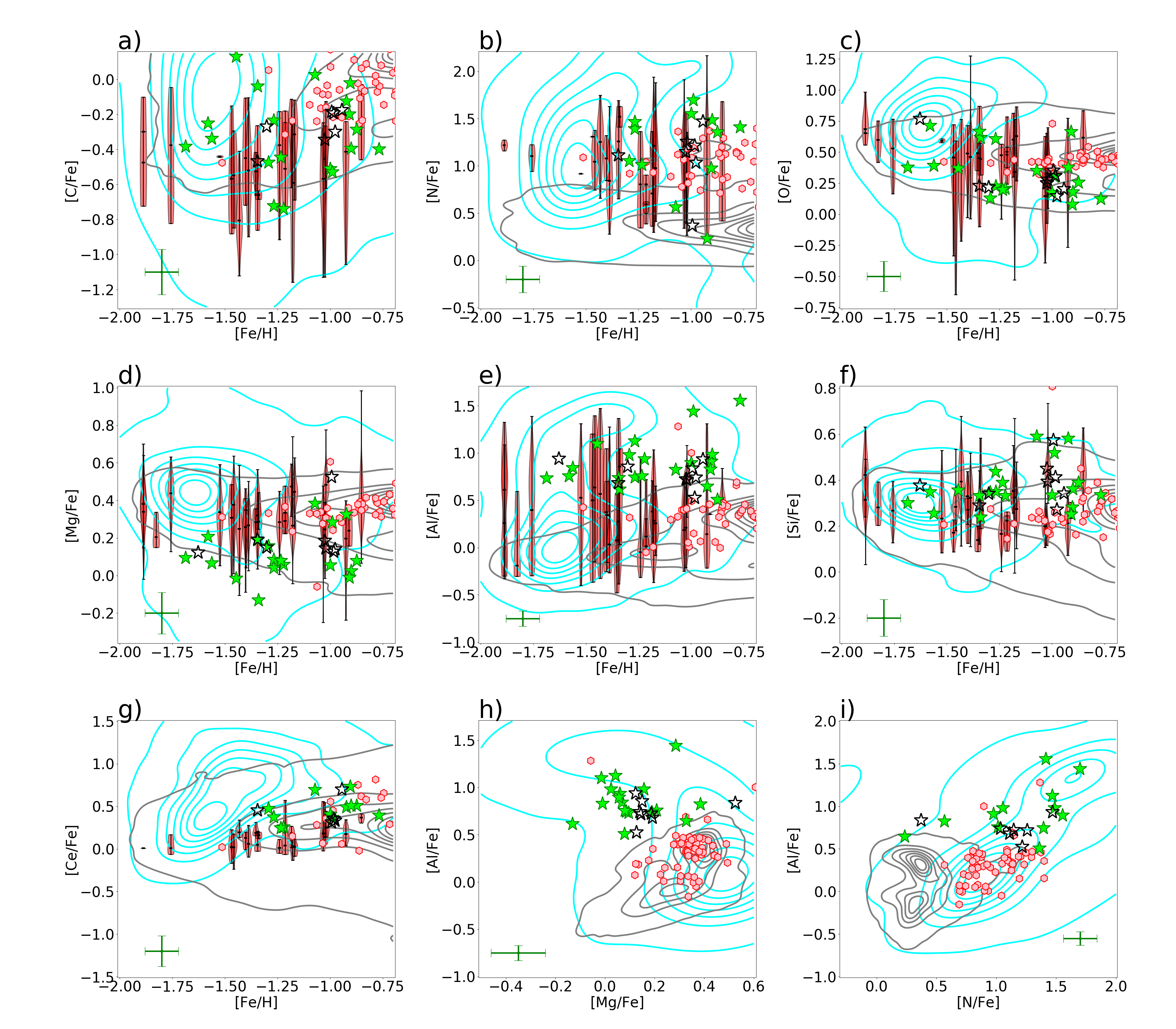}
		\caption{Kernel Density Estimation (KDE) of [X/Fe] (with X $=$ C, N, O, Mg, Al, Si, or Ce) with metallicity (a--g), and [Al/Fe] as a function of [Mg,N/Fe] ratios (i,h) for the APOGEE DR16 stars (grey contours) surviving the quality cuts discussed in Section \ref{section2} and corresponding to the chemical domain of the halo, and low-metallicity end of the bulge and disc, corrected by the offsets between pipelines (see Figure \ref{Figure7}). $\omega$ Centauri stars (cyan contours) from \citet{Szabolcs2020} are also plotted for comparison. The black and lime star symbols refer to Al-rich stars in the bulge sample and outside the bulge region, respectively. Our sample is compared to GCs (crimson violin representation, indicating with horizontal lines the median and limits of the distribution; \textit{left} to \textit{right}: M 53, NGC 6397, NGC 5466, M 55, M 22, M 79, NGC 6752, M 13, M 2, NGC 6544, M 3, M 54, M 10, NGC 6522, NGC 3201, NGC 6229, Pal 5, NGC 288, M 5, M 12, NGC 1851, NGC 362, M 4, NGC 2808, Pal 6, and M 107) from \citet{Szabolcs2020}, and a sample of selected NRS (pink hexagons) from \citet{Schiavon2017} with available APOGEE DR16 abundance ratios. The plotted error bars (green symbols) show the typical uncertainties of our sample.}
		\label{Figure1}
	\end{center}
\end{figure*}	
 
	\section{Elemental Abundance Analysis}
     \label{section3}
	
	Large amounts of aluminum were found in the atmospheres of 29 metal-poor field giant stars (not previously identified). Our sample comprises of 7 stars with orbital apocenter ($r_{apo}$) $\lessapprox$3.5 kpc (see Section \ref{orbit}), placing them well within the bulge region \citep[see, e.g.][]{Barbuy2018} and two halo interlopers located in the inner region of the Galaxy, which we refer here to as the bulge sample. The remaining stars in our sample are located within $\sim$ 13 kpc from the Galactic center. The light-elements (C, N), the $\alpha-$elements (O, Mg, Si), the Odd-Z element (Al), the iron-peak element (Fe), and the \textit{s}-process element (Ce) of these stars are compared to GC stars from \citet{Szabolcs2020}, and the local MW disc, halo and bulge population from APOGEE DR16 \citep{Ahumada2020}. Abundances have been determined from a $\chi^{2}$ minimization to synthesized spectra using the \texttt{BACCHUS} code.
	
	It is also important to note that there are no known GCs within an angular separation of 0.5 degrees for most of our stars, except for one (2M18035017$-$2552334), which is located $\sim1.2r_t$ from the GC Terzan 10. However, its [Fe/H] and RV deviate significantly from the nominal parameters of the cluster, so it cannot be considered a potential extra-tidal star candidate, as has been found recently in other bulge GCs (Fern\'andez-Trincado et al. 2020b, submitted). 
	
	Overall, we find that the elemental abundances of our sample is separated relatively cleanly from other stellar populations in different chemical dimensions, such as [Mg,N/Fe]--[Al/Fe] planes. Figure \ref{Figure1} reveal that our observed stars differ substantially from those of the so--called N-rich (or NRS) bulge population \citep[see, e.g.,][]{Schiavon2017} and from the rest of the entire MW/bulge sample, but exhibit a unique chemical signature similar to that of some GC 2G stars with higher [Al/Fe]. Also, their higher aluminum enrichment ([Al/Fe]$\gtrsim+0.5$) over a wide range of metallicities makes them unlikely to be associated to the typical Al enrichment seen in dwarf galaxies \citep[][]{Hasselquist2019}, unless they are part of partially dissipated GCs in dwarf galaxies (see Fern\'andez-Trincado et al. 2020c).
	
	\subsection{The iron-peak element: Fe}
	
	 Regarding the iron-peak element (Fe), we find that our sample spans a wide range of iron abundances, $-1.68 \lesssim$ [Fe/H] $\lesssim -0.76$ (see Figure \ref{Figure1}), thus suggesting that these stars were formed from different progenitors or in a unusual system like $\omega$ Cen (unless they are part of a binary system). It is interesting to note that our Al-rich stars toward the bulge region peak around [Fe/H]$\sim-1$, just as NRS stars -- located in bulge -- do \citep{Schiavon2017}, suggesting an association to them. It is then possible that we have identified a subfamily of that population which is strongly more enriched in aluminum, whilst the stars in our sample located outside the bulge region do no peak at any particular metallicity and are moderately more metal-poor. 

     \subsection{The $\alpha-$elements: O, Mg, and Si}
     
     Figure \ref{Figure1}(c) shows that the [O/Fe] abundance ratio of our sample is similar to that of the NRS, and MW/bulge stars at [Fe/H]$]\gtrsim-1.5$, but moderately enhanced toward the bulge region at metallicities as low as [Fe/H]$\lesssim-1.5$, and slightly enhanced for our sample outside the bulge region. In the same figures it is clear that [O/Fe] abundances of the Al-rich stars are, on average, consistent with the GCs population at similar metallicity, and have a large dispersion in [O/Fe] similar to that observed in GCs. 
     
      Regarding Magnesium (see Figure \ref{Figure1}(d)), most of our targets exhibit low levels of it ([Mg/Fe]$\lesssim+0.2$) and behave like accreted halo stars in the low-Mg sequence. At higher metallicity ([Fe/H]$\gtrsim-1.25$), a few stars show enhancement in [Mg/Fe] relative to the disc, providing evidence that these stars do not belong to canonical disk or halo of the MW, and neither to the bulge population, and have, on average, lower [Mg/Fe] than the NRS. It also indicates that the [Mg/Fe] levels and the star-to-star dispersion observed in our sample is consistent to that observed in GC stars and $\omega$ Cen stars. 
      
       The $\alpha-$element Si is found to be higher in our sample compared to the NRS and MW/bulge stars, but similar to those known to exist in $\omega$ Cen stars, and a few other Galactic GCs at similar metallicity. The slightly higher [Si/Fe] ($\sim+0.25$) of our stars indicate that the initial composition of the gas that formed them was slightly different from the NRS and MW stars, but similar to that of GC stars, providing further evidence that the Al-rich stars could be likely accreted stars from an progenitor whose chemical history was similar to that of known Galactic GCs.
      
       \subsection{The light-elements: C and N}

 		The outlying stars with high levels of Al enhancement are also enhanced in N, well above the Galactic levels ([N/Fe]$\gtrsim+0.8$) as can be noticed in Figure \ref{Figure1}(b;i), except for three stars at metallicities above [Fe/H]$\gtrsim-1.2$ that exhibit nitrogen abundance lower than [N/Fe]$\sim+0.5$, but clearly distinguishable from the canonical Galactic components in the [N/Fe]--[Al/Fe] plane. Furthermore, the Al-rich stars are slightly more enriched in nitrogen than the NRS, indicating that they do not necessarily share the same star-formation history, instead being similar to that of GCs environments. Interestingly, we can see a clear N-Al correlation in Figure \ref{Figure1} between our Al-rich stars toward the bulge region and the NRS, which is striking similar to that seen in Galactic GCs like $\omega$ Cen.  

		Regarding carbon, we find similar patterns as NRS and GC population at same metallicities, have slightly lower levels than the bulge, canonical disc and halo populations, extending down well below [C/Fe]$\lesssim-0.7$ dex.

       \subsection{The odd-Z elements: Al}   
 
 		Figure \ref{Figure1}(h) shows a well-distinguished clump of Al-rich stars, differing from the NRS, MW, and bulge stars, indicating that these stars are likely members of a distinct population confined to the inner part of the Galaxy (based in their orbital parameters, see subsection \ref{orbit}), possibly part of the innermost distribution of the stellar halo. There are also a few cases in the [Mg/Fe]--[Al/Fe] plane with large Al enhancement paired with low Mg, [Mg/Fe]$<0$, which is a signature of GC 2G stars \citep[see, e.g.,][]{Fernandez-Trincado2017, Lucey2019}. 

 		Overall, Figure \ref{Figure1}(h) displays the chemical-abundance pattern of GC stars from $\omega$ Cen \citep{Szabolcs2020}; this GC spans as wide a metallicity range as our observed Al-rich stars. The star-to-star scatter of the Al-rich stars for almost all the chemical species is similar to that typically found in $\omega$ Cen stars. The similarity between Al-rich stars and GC 2G stars can be further tested through examination of the [Mg, Al/Fe]--[Al/Fe] planes. Interestingly, we find that Al-rich stars exhibit similar characteristics to the NRS, i.e., they also show high [N/Fe], correlated with [Al/Fe] and anti-correlated with [C/Fe]. To first order, these branches strongly resemble those long known to exist in GCs, and are not present in the Galactic field. As far as we know, this is the first time the existence of clear Mg-Al anticorrelation and N-Al correlation among metal-poor stars with chemical anomalies toward the Galactic bulge is reported, and beyond the GC environments where typically these have been identified so far. This finding reinforces the unique nature of our sample, and confirms the complex and composite structure of the MW bulge.
 		 
        \subsection{The \textit{s}-process element: Ce}  
        
        The APOGEE DR16 [Ce/Fe] abundance ratio patterns \citep{Cunha2017} are shown for MW stars in Figure \ref{Figure1}(g) and stars in the bulge region in Figure \ref{Figure1}(g). Figure \ref{Figure1}(g) indicates that [Ce/Fe] is slightly enhanced than NRS, GC and MW stars, but fairly agree with GC stars \citep{Szabolcs2020} at similar metallicities, but there is currently no explanation for this unexpected tendency.
 
    	\subsection{Orbital Analysis}
    	\label{orbit}
	 
	  The model and distribution of orbital parameters are presented in Appendix \ref{orbits}. Overall, we find that all the stars in our sample have highly eccentric orbits ($e \gtrsim0.6$). Figure \ref{Figure3} shows the Galactic spatial distribution and the orbital elements for the Al-rich stars. As is clear from the figure, the majority of the stars are found to have radial ($r_{per}\lesssim3.5$ kpc) orbits in different configurations (e.g., retrograde, prograde, and P--R\footnote{We call prograde-retrograde (P-R) orbits to the ones that flip their sense from prograde to retrograde, or vice-versa, along its orbit.} orbits). The orbital elements also reveal that most of the Al-rich stars are currently located near their apocentric distances ($r_{apo}$), placing many of them within the Solar radius ($r_{gal}\lesssim8$ kpc) and inside of the bulge region ($r_{gal}\lesssim3$ kpc), with rather small excursions above the Galactic plane ($|Z|_{max} \lesssim$ 3 kpc). Several (11 out of 29 stars) of them have larger  (3 kpc $\lesssim |Z|_{max} \lesssim$ 25 kpc) vertical excursions, making them halo interlopers within the Solar radius. We conclude that these stars live in the bulge region and the inner stellar halo of the Galaxy, respectively.
	  
	 An interesting aspect of the Al-rich stars can be noticed in Figure \ref{Figure3}(h). Strikingly, this figure reveals that a disc control sample taken from APOGEE DR16 populates the top part of the $v_{\phi}$ vs. $v_{R}$ distribution, whilst the Al-rich stars fall in the bottom part, where the \textit{Gaia}-Sausage-Enceladus (GSE), Helmi Streams (Helmi St.) and the Arjuna$+$Sequoia$+$I'itoi (ASI) halo structures reside \citep[see, e.g.,][]{Naidu2020}, and which are dominated by metal-poor ($-3\lesssim$[Fe/H]$\lesssim-0.5$) stars. For comparison, we calculated the orbital solutions for Galactic GCs from \citet{Baumgardt2019}, and adopted the progenitor classification of \citet{Massari2019}. Figure \ref{Figure6} shows the characteristic orbital energy (($E_{max} + E_{min}$)/2) versus the orbital Jacobi constant ($E_{J}$) distribution of Galactic GCs and our Al-rich stars. This diagram reveals that the Al-rich stars populate a wide range of energies, similar to that of Galactic GCs with different origins, suggesting that the aforementioned have possibly emerged from different GCs.
	 
	 Are the Al-rich stars largerly the signatures of tidal disruption of accreted dwarf galaxies?. This possibility seems unlikely, given that the observed Al-rich stars exhibit higher [Al/Fe]$\gtrsim+0.5$, not observed in present-day dwarf galaxy stars \citep[see][and references therein]{Hasselquist2019}. Even though most of the Al-rich stars share similar orbital eccentricity, spatial distribution, and location in the $v_{\phi}$ vs. $v_{R}$ plane as those seen in high-$\alpha$ disc stars and in-situ halo stars, Metal-Weak Thick-Disc stars, Sagittarius-, Thamnos-, GSE-,  Helmi St.-, and ASI-structures \citep{Naidu2020}, their association seem unlikely, given the higher Al and N abundance observed. Our sample is also not part of the unclassified debris stars reported in \citet{Naidu2020}, as the Al-rich stars possess orbital eccentrities larger than those unclassified debris. However, since there are some GCs that appear to be possibly associated with some of the known mergers, in particular to GSE, the Helmi St., and Sequoia \citep[see, e.g.,][]{Massari2019}, then it is possible that some of our observed stars could be associated with those merger events through their partially dissipated GCs as highlighted in Figure \ref{Figure6}. 
	 
    All the above described chemical and dynamical properties suggest that the Al-rich stars could be the signatures of tidal disruption of accreted and likely formed in situ GCs. This population could have been dynamically ejected in different orbital configurations into the bulge region and the inner stellar halo from GC systems at similar metallicity, or possibly a massive system like $\omega$ Cen \citep[]{Meza2005, Majewski2012}. 	
	
    \section{Conclusions}
	 \label{section4}

	We report the detection of aluminum-enhanced ([Al/Fe]$\gtrsim+0.5$) giant stars toward the bulge and inner stellar halo of the Galaxy. The majority of these 29 stars have unique abundance patterns seen in accreted GCs, and dynamical properties consistent with the bulge and halo population. 
	
	The identification of such unusual stars toward the bulge region confirms the proposed composite nature of the MW bulge \citep[e.g.,][]{Recio-Blanco2017, Rojas-Arriagada2020, Queiroz2020b}. Our finding also reveals, for the first time, that there are many chemically anomalous metal-poor stars in the $v_{\phi}$ vs. $v_{R}$ plane residing in the same position where massive merger events (e.g., \textit{Gaia}-Sausage-Enceladus and Sequoia) have been identified. However, our study also reveals that some $\alpha-$elements (Si) and odd-Z elements (Al) appear to rule out any direct link with those known merger events, but some indirect association through GCs may still apply. 
	
	We conjecture that many of our Al-rich stars were possible former members of several stellar clusters with similar chemical enrichment as accreted-and-likely-formed-in-situ GCs. The peculiar chemical signature of these objects, makes them excellent candidates to trace the fossil relics population of early MW, which became part of the general stellar population.
	 	
	\begin{acknowledgements}  
	The author is grateful for the enlightening feedback from the anonymous referee. J.G.F-T is supported by FONDECYT No. 3180210. T.C.B. acknowledges partial support for this work from grant PHY 14-30152: Physics Frontier Center / JINA Center for the Evolution of the Elements (JINA-CEE), awarded by the US National Science Foundation. D.M. is supported by the BASAL Center for Astrophysics and Associated Technologies (CATA) through grant AFB 170002, and by project FONDECYT Regular No. 1170121. B.T. gratefully acknowledges support from the National Natural Science Foundation of China under grant No. U1931102 and support from the hundred-talent project of Sun Yat-sen University. S.V. gratefully acknowledeges the support provided by Fondecyt regular n. 1170518. D.G. gratefully acknowledges support from the Chilean Centro de Excelencia en Astrof\'isica y Tecnolog\'ias Afines (CATA) BASAL grant AFB-170002. D.G. also acknowledges financial support from the Direcci\'on de Investigaci\'on y Desarrollo de la Universidad de La Serena through the Programa de Incentivo a la Investigaci\'on de Acad\'emicos (PIA-DIDULS). A.P-V acknowledges FAPESP for the postdoctoral fellowship No. 2017/15893-1 and the DGAPA-PAPIIT grant IG100319. The SDSS-IV/APOGEE-2 survey made this study possible.
	\end{acknowledgements}
	

\begin{appendix}
	
\section{Sample selection}
\label{sampleobservations}

APOGEE DR16 includes data taken from both the Northern and Southern Hemisphere using the APOGEE spectrographs \citep{Wilson2019} mounted on the 2.5m Sloan Foundation telescope \citep{Gunn2006} at Apache Point Observatory in New Mexico (APOGEE-2N: North, APO), and on the 2.5m Ir\'en\'ee du Pont telescope \citep{Bowen1973} at Las Campanas Observatory (APOGEE-2S: South, LCO) in Chile. For details regarding the APOGEE atmospheric parameters analysis we direct the reader to the APOGEE Stellar Parameter and Chemical Abundances  pipeline \citep[\texttt{ASPCAP}:][]{GarciaPerez2016a}, while for details about the grid of synthetic spectra and errors see \citet{Holtzman2018}. We refer the reader to \citet{Nidever2015} for further details about the data reduction pipeline for APOGEE. The model grids for APOGEE DR16 are based on a complete set of \texttt{MARCS} \citep{Gustafsson2008} stellar atmospheres, which are now extended to effective temperatures ($T_{\rm eff}$) as low as 3200 K.

The parent sample upon which this work is based is defined as follows. Metal-poor stars were selected from the DR16 catalogue\footnote{\url{https://www.sdss.org/dr16/irspec/spectro_data/}} that match the following criteria: (\textit{i}) $-2 \lesssim$ [Fe/H]$\lesssim -0.7$; imposing a lower limit on metallicity, i.e., [Fe/H] $>$ $-2$, allows for the inclusion of metal-poor stars with high-quality spectra with reliable parameters and chemical abundances, while requiring stars with metallicity below [Fe/H]$=-$0.7 minimizes the presence of disc stars; (\textit{ii}) stars with S/N $>$ 60 were selected, to ensure that we are selecting; spectra that have well-known uncertainties in their stellar parameters and chemical abundances, and to remove stars with lower quality spectra; (\textit{iii}) 3200 K $ < T_{\rm eff} < $ 5500 K, this temperature range ensures that stellar parameters are reliably and consistently determined, and maximizes the overall quality of the abundances considered; (\textit{iv}) the estimated surface gravity ($\log$ \textit{g}) must be less than 3.6, to ensure inclusion of giant and subgiant stars, which have more accurate \texttt{ASPCAP}-parameters than the dwarf stars with $\log$ \textit{g} $>$3.6 (due to the lack of asteroseismic surface gravities for dwarfs, only stars with  $\log$ \textit{g} $<$ 3.6 have calibrated surface gravities); (\textit{v}) \texttt{ASPCAPFLAG} $==$ \texttt{0}; this cut ensures that there were no major flagged issues, i.e., low signal-to-noise, poor synthetic spectral fit, stellar parameters near grid boundaries, among others; and (\textit{vi}) stars belonging to Galactic GCs from \citet{Szabolcs2020}, anomalous stars from \citet{Martell2016, Schiavon2017, Fernandez-Trincado2016, Fernandez-Trincado2017, Fernandez-Trincado2019a, Fernandez-Trincado2019b, Fernandez-Trincado2019c, Fernandez-Trincado2019d}, and stars with carbon abundances [C/Fe] $\gtrsim +0.15$ (to ensure the low carbon abundances typical of GC stars) were excluded from the sample. 

The final selected sample amounts to a total of 13,338 stars with high-quality parameters. We search for aluminum-enriched stars by carrying out a semi-independent stellar-abundance analysis of Al I lines by using the \texttt{BACCHUS}\footnote{BACCHUS: {\bf B}russels {\bf A}utomatic {\bf C}ode for {\bf C}haracterizing {\bf H}igh acc{\bf U}racy {\bf S}pectra} code \citep{Masseron2016}, following the method described by \citet{Fernandez-Trincado2019b}. 	
	
\section{Basic parameters}	
	
	In Table \ref{Table1} we provide the abundances of C, N, O, Mg, Al, Si, Fe, and Ce for the 29 stars analysed in this work. The abundances were determined by making a careful line selection, and a line-by-line analysis in Local Thermodynamic Equilibrium (LTE) with the \texttt{BACCHUS} code \citep{Masseron2016}. The T$_{\rm eff}$, $\log$ \textit{g}, and overall metallicity ([M/H]) have been fixed, and were selected from uncalibrated \texttt{ASPCAP} T$_{\rm eff}$, $\log$ \textit{g}, and [M/H] from DR16. In addition, lines were rejected if they were found to not be well-reproduced by the synthesis, or if they were flagged as problematic by the \texttt{BACCHUS} pipeline. None of the 29 Al-rich star candidates exhibit particularly strong variability in its radial velocity (V$_{\rm scatter} \lesssim 0.5$ km s$^{-1}$) over the period ($\lesssim 6$ months) of the APOGEE observations, therefore with the existing data the observed abundace anomalies could be not explained by channels invoking binary mass transfer or stellar variability. The uncertainties in the elemental abundances have been computed by adding the typical sensitivities of the abundance and the internal error in quadrature, in the same manner, as described in \citet{Fernandez-Trincado2019a, Fernandez-Trincado2019b, Fernandez-Trincado2019c, Fernandez-Trincado2019d}. The typical uncertainties are shown with two crossed green error bars (one per axis plotted) in Figure \ref{Figure1}.
	
	\begin{sidewaystable*}
	\begin{small}
		\setlength{\tabcolsep}{2.85mm}  
		\caption{Basic parameters of the Al-rich stars.}
		\centering
		\begin{tabular}{|c|ccccccccccccccc|}
			\hline
APOGEE\_ID &      T$_{\rm eff}$   & $\log$ \textit{g} & [M/H] & $\mu_{t}$  & ${\rm V_{scatter}}$ & S/N  & \#Visits & [C/Fe] & [N/Fe]  & [O/Fe]  & [Mg/Fe] & [Al/Fe]  & [Si/Fe] & [Fe/H] & [Ce/Fe] \\
 &      K   &  & & km s$^{-1}$  &  km s$^{-1}$ & pixel$^{-1}$  &  &  &  &   & &  & &  & \\
\hline
2M16035902$+$0703353 & 5095 & 2.71 & $-$1.34 & 0.6 & 0.13 & 147 & 6 & $-$0.03 & ...  & 0.66 & $ $0.18 & 0.73 & 0.33  & $-$1.34  & ...  \\
2M17343682$-$2654206 & 4511 & 1.98 & $-$0.98 & 1.8 & 0.11 &  80 & 2 & $ $...  & 1.25 & 0.27 & $ $0.14 & 0.71 & 0.39  & $-$1.02  & ...  \\
2M17254298$-$2106351 & 4740 & 1.89 & $-$1.52 & 1.1 & 0.09 &  74 & 2 & $ $...  & ...  & 0.76 & $ $0.12 & 0.94 & 0.37  & $-$1.62  & ...  \\
2M21372238$+$1244305 & 5062 & 2.86 & $-$0.93 & 0.9 & 0.40 & 127 & 4 & $-$0.19 & ...  & 0.66 & $-$0.01 & 0.82 & 0.25  & $-$0.90  & ...  \\
2M17461523$-$3407127 & 4559 & 1.93 & $-$1.22 & 1.6 & ...  &  84 & 1 & $-$0.26 & ...  & 0.21 & $ $0.14 & 0.85 & 0.34  & $-$1.30  & ...  \\
2M17313399$-$2640506 & 3792 & 0.62 & $-$0.93 & 2.7 & 0.40 & 232 & 2 & $-$0.19 & 1.21 & 0.30 & $ $0.12 & 0.52 & 0.41  & $-$0.98  & 0.35 \\
2M19010991$-$2121082 & 4368 & 1.39 & $-$1.33 & 1.7 & 0.06 & 346 & 3 & $-$0.46 & 1.11 & 0.22 & $ $0.19 & 0.68 & 0.29  & $-$1.34  & 0.45 \\
2M16443329$+$3624504 & 4816 & 2.06 & $-$1.36 & 1.3 & 0.04 & 164 & 3 & $ $0.13 & ...  & 0.37 & $-$0.01 & 1.10 & 0.35  & $-$1.44  & ...  \\
2M15535622$+$2237419 & 4284 & 1.27 & $-$1.26 & 1.8 & 0.15 & 177 & 2 & $-$0.47 & 1.05 & 0.12 & $ $0.15 & 0.98 & 0.34  & $-$1.29  & 0.47 \\
2M15053060$+$4211466 & 4689 & 1.72 & $-$1.52 & 1.4 & 0.09 & 122 & 4 & $-$0.33 & ...  & 0.39 & $ $0.06 & 0.84 & 0.25  & $-$1.56  & ...  \\
2M16282719$-$0050428 & 4433 & 1.95 & $-$1.05 & 1.9 & 0.21 & 128 & 6 & $ $0.02 & 0.56 & 0.35 & $ $0.38 & 0.82 & 0.59  & $-$1.07  & 0.69 \\
2M18041915$-$3703227 & 3808 & 0.33 & $-$1.07 & 1.8 & 0.02 & 239 & 2 & $-$0.18 & 0.37 & 0.35 & $ $0.52 & 0.83 & 0.57  & $-$0.99  & 0.31 \\
2M15532396$-$3509322 & 4766 & 2.12 & $-$1.53 & 0.8 & ...  &  89 & 1 & $-$0.25 & ...  & 0.71 & $ $0.20 & 0.75 & 0.34  & $-$1.58  & ...  \\
2M17425767$-$2803437 & 4569 & 1.96 & $-$0.85 & 2.9 & 0.39 &  92 & 2 & $-$0.17 & 1.46 & 0.20 & $ $...  & 0.93 & 0.34  & $-$0.94  & 0.69 \\
2M18490114$-$3042332 & 4426 & 1.79 & $-$0.96 & 2.1 & ...  & 159 & 1 & $-$0.29 & 1.03 & 0.14 & $ $0.13 & 0.74 & 0.27  & $-$0.97  & 0.33 \\
2M16055435$+$3003432 & 4469 & 1.34 & $-$1.65 & 1.8 & 0.08 & 185 & 2 & $-$0.38 & ...  & 0.37 & $ $0.09 & 0.73 & 0.30  & $-$1.68  & ...  \\
2M17254242$-$1331201 & 4485 & 1.99 & $-$0.97 & 2.1 & 0.08 & 256 & 3 & $-$0.51 & 1.55 & 0.17 & $ $0.05 & 0.89 & 0.33  & $-$0.99  & 0.40 \\
2M18524618$-$2520372 & 4518 & 2.07 & $-$1.02 & 1.5 & ...  &  70 & 1 & $-$0.34 & 1.14 & 0.24 & $ $0.18 & 0.72 & 0.44  & $-$1.02  & ...  \\
2M14063545$+$2734116 & 4859 & 2.17 & $-$1.32 & 1.8 & 0.28 &  74 & 6 & $ $...  & ...  & 0.61 & $-$0.13 & 0.61 & 0.23  & $-$1.34  & ...  \\
2M16102218$-$2158014 & 4342 & 1.55 & $-$1.24 & 2.1 & ...  & 326 & 1 & $-$0.72 & 1.39 & 0.22 & $ $0.08 & 0.74 & 0.35  & $-$1.26  & 0.37 \\
2M02593571$-$1731458 & 4472 & 2.09 & $-$0.96 & 1.3 & 0.46 &  87 & 3 & $-$0.12 & 0.23 & 0.37 & $ $0.32 & 0.64 & 0.58  & $-$0.92  & 0.49 \\
2M22015198$-$1141479 & 4269 & 1.57 & $-$1.25 & 1.8 & 0.15 & 279 & 3 & $-$0.44 & 1.01 & 0.19 & $ $0.07 & 0.75 & 0.39  & $-$1.23  & 0.24 \\
2M16112935$-$2000538 & 4871 & 2.35 & $-$1.20 & 1.1 & ...  & 117 & 1 & $-$0.23 & 1.46 & 0.60 & $ $0.04 & 1.12 & 0.43  & $-$1.26  & ...  \\
2M06572697$+$5543115 & 4776 & 2.22 & $-$1.21 & 1.2 & 0.02 & 446 & 3 & $-$0.73 & ...  & 0.20 & $ $0.05 & 0.93 & 0.33  & $-$1.22  & 0.25 \\
2M17431262$-$2843363 & 4149 & 0.16 & $-$0.72 & 2.0 & ...  & 115 & 1 & $-$0.39 & 1.41 & 0.12 & $ $...  & 1.55 & 0.33  & $-$0.76  & 0.39 \\
2M17475763$-$2913591 & 3837 & 0.06 & $-$1.04 & 2.0 & ...  & 100 & 1 & $-$0.52 & 1.69 & 0.30 & $ $0.28 & 1.44 & 0.51  & $-$0.98  & 0.36 \\
2M17480799$-$2712083 & 3913 & 1.11 & $-$0.74 & 2.9 & ...  &  90 & 1 & $-$0.28 & 1.35 & 0.25 & $ $0.07 & 0.50 & 0.38  & $-$0.87  & 0.50 \\
2M17484544$-$2915516 & 4241 & 1.53 & $-$0.82 & 2.9 & 0.02 &  81 & 2 & $-$0.01 & 0.97 & 0.08 & $ $...  & 0.91 & 0.28  & $-$0.90  & 0.73 \\
2M18035017$-$2552334 & 4190 & 1.59 & $-$0.88 & 2.3 & ...  & 137 & 1 & $-$0.39 & 1.48 & 0.17 & $ $0.02 & 0.98 & 0.36  & $-$0.90  & 0.50 \\
			\hline
		\end{tabular}  \label{Table1}
	\end{small}
\end{sidewaystable*}

\section{Galactic orbits}	
\label{orbits}

The basic kinematic parameters for the 29 stars and their respective orbital elements are listed in Table \ref{Table2}. Since the true Galactic potential is not accurately known in the inner Galaxy, and the results of our simulations may depend significantly on the assumed parameters, we also run the simulations with a slow and fast rotating bar, i.e., 31 km s$^{-1}$ and 51 km s$^{-1}$, respectively, and see how it affects the derived orbital parameters of the stars. The value inside parentheses in Table \ref{Table2} show the effects in the orbital elements by adopting different angular velocity for the Galactic bar. The individual variations are within typical errors of the ensemble of orbits with $\Omega_{\rm bar} = 41$ km s$^{-1}$, which do not change the overall results from our sample.

\begin{sidewaystable*}
	\begin{tiny}
		\setlength{\tabcolsep}{1.85mm}  
		\caption{Orbital elements of the Al-rich stars.}
		\centering
		\begin{tabular}{|c|ccccccccccc|}
			\hline
			APOGEE\_ID  &  \texttt{RUWE} & $d_{\odot} \pm \Delta $ &   RV & $\mu_{\alpha}\cos{}(\delta) \pm \Delta$     &    $\mu_{\delta} \pm \Delta$  &   $r_{peri}$ &  $r_{apo}$   &  |Z|$_{\rm max}$   & eccentricity & $L_{Z}^{min}$ &    $L_{Z}^{max}$      \\
			&  & kpc &   km s$^{-1}$ & mas yr$^{-1}$  &   mas yr$^{-1}$  &   kpc &  kpc   &  kpc  &  & $\times 10^{2}$ km s$^{-1}$ kpc &   $\times 10^{2}$ km s$^{-1}$ kpc     \\ 
			\hline
			2M16035902$+$0703353 & 1.06 &  3.7$\pm$0.4  &     33.5 & $-$10.28$\pm$0.04 &  $-$6.17$\pm$0.03 &  0.76$\pm$0.47 (0.58) &  5.92$\pm$0.33 (0.49) &  4.55$\pm$0.51 (0.53) & 0.77$\pm$0.12 (0.12) &  $-$44.0$\pm$14.0(14.40) &  $-$19.5$\pm$14.5(16.59) \\
			2M17343682$-$2654206 & 0.96 &  6.5$\pm$0.9  &     61.9 & $-$10.14$\pm$0.18 &  $-$1.71$\pm$0.15 &  0.03$\pm$0.16 (0.26) &  2.24$\pm$0.48 (0.65) &  2.20$\pm$0.23 (0.23) & 0.96$\pm$0.08 (0.11) &  $-$11.0$\pm$10.5(14.09) &   $ $5.0$\pm$14.0(13.01) \\
			2M17254298$-$2106351 & 0.96 &  7.0$\pm$0.9  &    146.0 &  $-$7.65$\pm$0.08 &  $-$3.63$\pm$0.06 &  0.05$\pm$0.21 (0.19) &  2.85$\pm$0.67 (0.67) &  2.14$\pm$0.13 (0.22) & 0.96$\pm$0.10 (0.09) &  $-$25.0$\pm$ 9.5(10.46) &   $ $6.0$\pm$13.5( 9.03) \\
			2M21372238$+$1244305 & 0.90 &  1.9$\pm$0.1  &  $-$85.7 &  $-$7.59$\pm$0.03 & $-$27.14$\pm$0.04 &  1.52$\pm$0.08 (0.20) & 10.28$\pm$0.37 (0.45) &  3.48$\pm$0.33 (0.63) & 0.74$\pm$0.02 (0.03) &  $-$72.0$\pm$ 3.5( 4.94) &  $-$57.0$\pm$ 2.0( 7.84) \\
			2M17461523$-$3407127 & 0.87 &  5.2$\pm$0.7  &     97.9 &  $-$6.02$\pm$0.07 &  $-$7.09$\pm$0.05 &  0.03$\pm$0.05 (0.14) &  3.19$\pm$0.69 (0.77) &  1.17$\pm$0.38 (0.51) & 0.97$\pm$0.03 (0.05) &  $-$21.0$\pm$ 7.0(13.95) &  $ $10.0$\pm$ 9.0( 9.67) \\
			2M17313399$-$2640506 & 0.88 &  6.5$\pm$0.8  &     74.9 &  $-$4.48$\pm$0.22 &  $-$7.18$\pm$0.15 &  0.02$\pm$0.01 (0.07) &  1.67$\pm$0.73 (0.70) &  0.62$\pm$0.11 (0.17) & 0.97$\pm$0.01 (0.04) &  $-$18.0$\pm$11.0(11.83) &   $ $5.0$\pm$ 3.5( 3.83) \\
			2M19010991$-$2121082 & 0.99 &  5.5$\pm$0.7  &   $-$9.7 &  $-$3.97$\pm$0.08 & $-$12.41$\pm$0.06 &  0.03$\pm$0.04 (0.08) &  3.93$\pm$0.30 (0.35) &  2.47$\pm$0.40 (0.52) & 0.98$\pm$0.02 (0.03) &   $-$6.0$\pm$ 8.5( 9.55) &  $ $17.0$\pm$ 9.5( 9.05) \\
			2M16443329$+$3624504 & 1.12 &  6.0$\pm$0.7  & $-$250.3 &  $-$3.28$\pm$0.02 &  $-$2.50$\pm$0.03 &  0.49$\pm$0.17 (0.21) &  7.57$\pm$0.40 (0.71) &  6.94$\pm$0.21 (0.95) & 0.87$\pm$0.03 (0.04) &  $ $14.0$\pm$ 6.0(10.90) &  $ $42.0$\pm$ 9.5(12.24) \\
			2M15535622$+$2237419 & 1.11 &  6.1$\pm$0.7  &  $-$96.9 &  $-$3.13$\pm$0.04 &  $-$2.46$\pm$0.04 &  1.03$\pm$0.36 (0.41) &  7.33$\pm$0.57 (0.40) &  5.59$\pm$0.57 (0.63) & 0.76$\pm$0.07 (0.08) &  $-$60.0$\pm$15.5(12.93) &  $-$29.0$\pm$11.0(11.79) \\
			2M15053060$+$4211466 & 0.98 & 10.1$\pm$1.2  & $-$228.2 &  $-$2.94$\pm$0.02 &  $-$1.50$\pm$0.03 &  0.44$\pm$0.39 (0.32) & 12.25$\pm$0.53 (0.88) & 12.32$\pm$0.59 (0.68) & 0.92$\pm$0.05 (0.04) &  $ $10.0$\pm$19.0(15.83) &  $ $23.0$\pm$17.5(15.33) \\
			2M16282719$-$0050428 & 1.06 &  7.9$\pm$1.0  &     12.5 &  $-$2.69$\pm$0.07 &  $-$4.51$\pm$0.03 &  0.31$\pm$0.21 (0.25) &  5.18$\pm$0.76 (0.64) &  4.10$\pm$0.48 (0.53) & 0.88$\pm$0.07 (0.09) &  $-$35.0$\pm$12.0( 9.27) &   $-$4.0$\pm$ 5.5( 6.95) \\
			2M18041915$-$3703227 & 1.01 & 10.8$\pm$0.4  &  $-$11.0 &  $-$2.61$\pm$0.08 &  $-$5.73$\pm$0.06 &  0.23$\pm$0.13 (0.25) &  3.42$\pm$0.35 (0.47) &  1.79$\pm$0.27 (0.22) & 0.87$\pm$0.06 (0.10) &  $-$35.0$\pm$ 5.0(10.85) &   $-$5.0$\pm$ 4.5( 8.41) \\
			2M15532396$-$3509322 & 1.03 &  5.6$\pm$0.7  &    208.1 &  $-$1.65$\pm$0.05 &  $-$7.39$\pm$0.03 &  0.04$\pm$0.03 (0.42) &  6.70$\pm$0.58 (1.00) &  2.58$\pm$0.32 (0.44) & 0.98$\pm$0.00 (0.15) &  $-$11.0$\pm$11.0(18.76) &  $ $30.0$\pm$12.5(12.58) \\
			2M17425767$-$2803437 & 1.16 &  6.5$\pm$1.2  &  $-$57.4 &  $-$1.33$\pm$0.47 &  $-$5.05$\pm$0.34 &  0.17$\pm$0.22 (0.37) &  1.68$\pm$1.04 (1.00) &  0.24$\pm$0.07 (0.10) & 0.79$\pm$0.11 (0.13) &  $-$22.0$\pm$23.0(21.86) &   $-$3.0$\pm$ 6.5(11.14) \\
			2M18490114$-$3042332 & 1.10 &  5.3$\pm$0.6  & $-$136.9 &  $-$1.15$\pm$0.03 &  $-$7.29$\pm$0.03 &  0.04$\pm$0.23 (0.22) &  3.59$\pm$0.61 (0.58) &  2.17$\pm$0.28 (0.26) & 0.97$\pm$0.08 (0.08) &  $-$30.0$\pm$16.5(13.22) &   $ $3.0$\pm$16.0(12.72) \\
			2M16055435$+$3003432 & 0.98 &  6.1$\pm$0.7  & $-$259.0 &  $-$0.44$\pm$0.04 &  $ $3.07$\pm$0.05 &  1.09$\pm$0.35 (0.36) & 13.81$\pm$0.90 (1.13) & 14.06$\pm$1.09 (1.29) & 0.85$\pm$0.05 (0.05) &  $-$33.0$\pm$12.0(11.74) &  $-$31.0$\pm$10.5(11.17) \\
			2M17254242$-$1331201 & 0.90 &  4.5$\pm$0.5  &   $-$7.9 &  $-$0.40$\pm$0.05 &  $-$5.35$\pm$0.04 &  1.06$\pm$0.13 (0.45) &  4.04$\pm$0.35 (0.35) &  1.22$\pm$0.16 (0.20) & 0.57$\pm$0.03 (0.10) &  $-$68.0$\pm$15.5(11.73) &  $-$30.0$\pm$ 4.5(11.90) \\
			2M18524618$-$2520372 & 1.00 &  6.6$\pm$0.8  &     19.1 &     0.06$\pm$0.07 &  $-$5.25$\pm$0.06 &  0.35$\pm$0.32 (0.28) &  2.66$\pm$0.36 (0.31) &  1.55$\pm$0.16 (0.18) & 0.76$\pm$0.18 (0.15) &  $-$31.0$\pm$11.5(10.93) &   $-$8.0$\pm$10.5(10.89) \\
			2M14063545$+$2734116 & 0.91 & 11.0$\pm$1.1  & $-$160.0 &     0.23$\pm$0.05 &  $-$3.17$\pm$0.05 &  2.20$\pm$0.19 (0.21) & 16.79$\pm$1.42 (1.33) & 15.17$\pm$1.55 (1.50) & 0.77$\pm$0.03 (0.03) &  $-$85.0$\pm$ 5.0( 6.11) &  $-$79.0$\pm$ 5.0( 5.99) \\
			2M16102218$-$2158014 & 1.13 &  3.0$\pm$0.4  &    153.2 &     0.85$\pm$0.10 &  $-$4.73$\pm$0.07 &  1.60$\pm$0.70 (0.51) &  9.19$\pm$0.85 (0.69) &  2.31$\pm$0.13 (0.40) & 0.67$\pm$0.08 (0.06) & $-$111.0$\pm$13.5(12.07) &  $-$55.0$\pm$22.0(17.38) \\
			2M02593571$-$1731458 & 1.00 &  7.5$\pm$0.6  &    327.7 &     2.39$\pm$0.04 &  $-$1.50$\pm$0.05 &  3.24$\pm$0.17 (0.16) & 30.91$\pm$0.74 (0.71) & 25.88$\pm$0.91 (0.99) & 0.81$\pm$0.01 (0.01) & $-$137.0$\pm$ 5.0( 5.10) & $-$135.0$\pm$ 5.5( 5.39) \\
			2M22015198$-$1141479 & 0.98 &  5.4$\pm$0.6  &  $-$76.4 &     3.27$\pm$0.07 &  $-$3.02$\pm$0.07 &  1.27$\pm$0.27 (0.41) &  8.35$\pm$0.25 (0.37) &  5.21$\pm$0.84 (0.82) & 0.73$\pm$0.04 (0.07) &  $-$70.0$\pm$15.0(14.83) &  $-$37.0$\pm$ 8.0(12.60) \\
			2M16112935$-$2000538 & 1.08 &  5.4$\pm$1.0  &     44.0 &     3.99$\pm$0.08 &  $-$9.61$\pm$0.05 &  0.75$\pm$0.57 (0.63) &  7.48$\pm$0.61 (0.93) &  5.02$\pm$1.48 (1.73) & 0.81$\pm$0.13 (0.13) &  $-$41.0$\pm$23.0(22.97) &  $-$23.0$\pm$21.5(22.53) \\
			2M06572697$+$5543115 & 0.96 &  2.1$\pm$0.2  & $-$239.3 &    10.03$\pm$0.06 & $-$14.92$\pm$0.05 &  0.55$\pm$0.23 (0.26) & 14.76$\pm$0.96 (0.84) &  2.42$\pm$0.65 (0.86) & 0.92$\pm$0.03 (0.03) &  $-$41.0$\pm$ 8.0(15.33) &  $-$23.0$\pm$12.5(18.93) \\
			2M17431262$-$2843363 & ...  &  ...          &      1.2 &    ...            &    ...            &  ...                  & ...                   &  ...                  & ...                  &  ...                     &  ...                     \\
			2M17475763$-$2913591 & ...  &  ...          & $-$114.1 &    ...            &    ...            &  ...                  & ...                   &  ...                  & ...                  &  ...                     &  ...                     \\
			2M17480799$-$2712083 & ...  &  ...          &     24.1 &    ...            &    ...            &  ...                  & ...                   &  ...                  & ...                  &  ...                     &  ...                     \\
			2M17484544$-$2915516 & ...  &  ...          & $-$218.7 &    ...            &    ...            &  ...                  & ...                   &  ...                  & ...                  &  ...                     &  ...                     \\
			2M18035017$-$2552334 & ...  &  ...          & $-$153.8 &    ...            &    ...            &  ...                  & ...                   &  ...                  & ...                  &  ...                     &  ...                     \\
			\hline
		\end{tabular}  \label{Table2}
	\end{tiny}
\end{sidewaystable*}
	
We used the \texttt{GravPot16}\footnote{\url{https://gravpot.utinam.cnrs.fr}} model to study the orbital history of our sample. The orbital integration scheme follows the methodology introduced in \citet{Fernandez-Trincado2020}, except for the angular velocity of the bar, for which we adopted $\Omega_{\rm bar} = 41$ km s$^{-1}$ kpc$^{-1}$ \citep[see, e.g.,][]{Bovy2019}. The model has been rescaled to the Sun's Galactocentric distance, 8 kpc, and the velocity of the Local Standard of Rest, $\nu_{LSR} = 244.5$ km s$^{-1}$.

\begin{figure*}	
	\begin{center}
		\includegraphics[width=190mm]{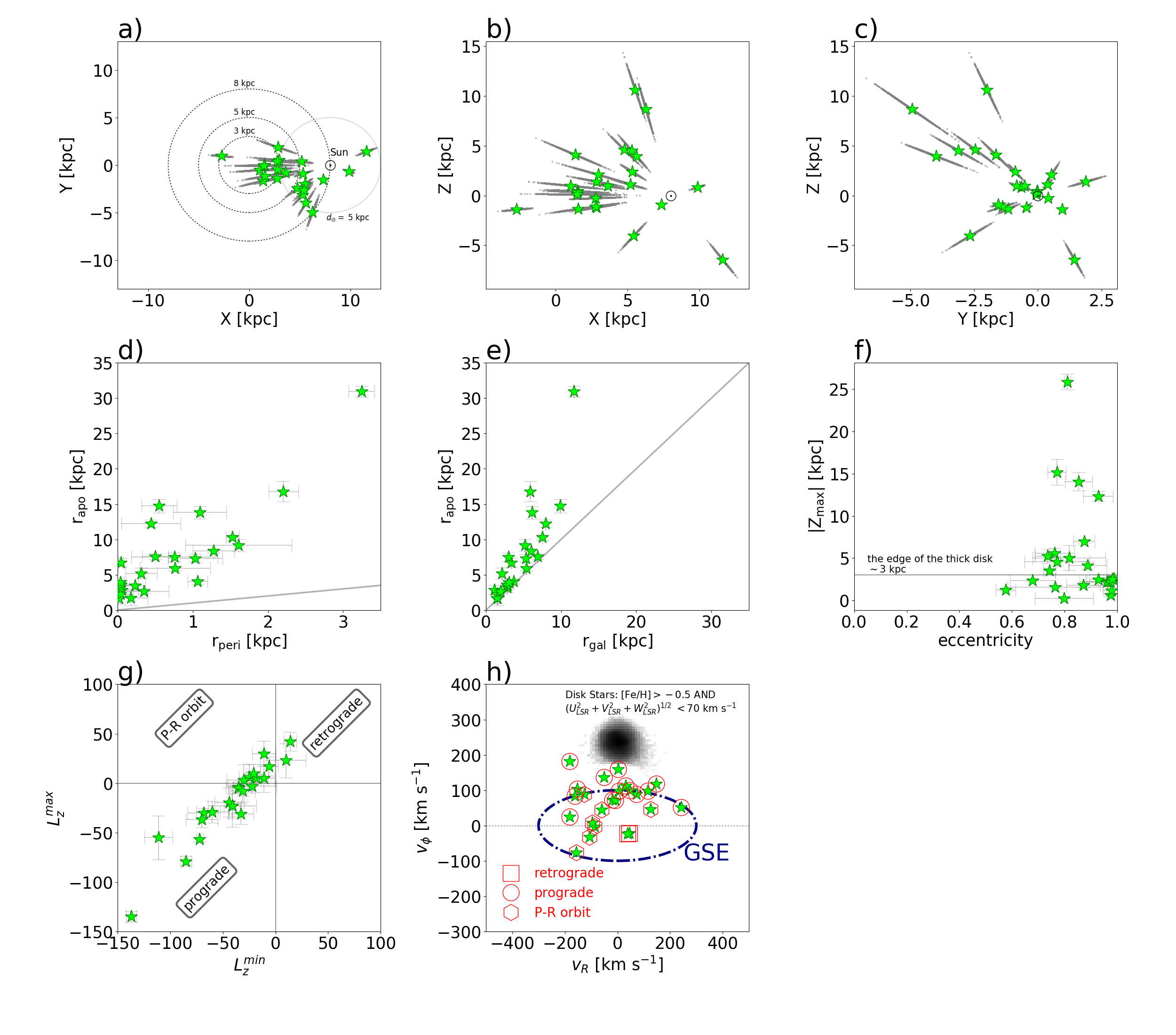}
		\caption{Panels (a) to (c) show the Galactic location of the Al-rich stars (lime star symbols). The aligned grey points show the results of a thousand Monte Carlo resampling incorporating measurement errors. The orbital elements are shown in panels (d) to (g). The straight lines in panels (d) and (e) indicate the one-to-one line. For panel (d), a star on this line would have a circular orbit, and in panel (e), a star on this line is being observed at apocentre. The horizontal line in panel (f) indicates the approximate edge of the thick disc according to \citet{Carollo2010}. The type of orbital configuration of each star is shown in panel (g). The median of the orbital parameters for each star is shown with their respective 16$^{th}$ and 84$^{th}$ percentiles (error bars) obtained for the model with the bar having $\Omega_{\rm bar} = 41$ km s$^{-1}$ kpc$^{-1}$. The distribution of the velocity components $v_{\rm \theta}$ vs. $v_{\rm R}$ for the studied Al-rich stars is shown in panel (h). The blue dashed line represents the approximate region for stars associated with \textit{Gaia}-Sausage in $v_{\rm \theta}$ vs. $v_{\rm R}$ space based on \citet{Belokurov2018}. In addition, a 2-D heat map (grey) of the densely distributed disc stars from APOGEE DR16 is shown for comparison. The unfilled red symbols in panel (h) mark the type of orbital configuration.}
		\label{Figure3}
	\end{center}
\end{figure*}

For each star, we integrated an ensemble of orbits over a 2 Gyr timespan. The uncertainties in the input data were randomly propagated as 1$\sigma$ variations in a Gaussian Monte Carlo re-sampling \citep[e.g.,][]{Fernandez-Trincado2020}. The resulting values were estimated as the median of a thousand realizations, and the error distributions by finding the 16$^{\rm th}$ and 84$^{\rm th}$ percentiles. In addition, we also obtained the change in the orbital elements when the bar angular velocity is varied by $\pm10$ km s$^{-1}$ kpc (values inside parentheses in Table \ref{Table2}). Figure \ref{Figure3} follows the orbital elements of our sample.  

We use the APOGEE DR16 distances deduced from the \texttt{StarHorse} code \citep{Queiroz2020a, Queiroz2020b}, radial velocities from the APOGEE survey, and absolute proper motions from the \textit{Gaia} DR2 catalog \citep{Gaia2018}. Only sources with 'good' astrometry were considered in our orbit computations, i.e., stars with Renormalized Unit Weight Error, \texttt{RUWE} $<$ 1.4 \citep[see, e.g.,][]{Lindegren2018}. A total of 24 out of 29 stars in our sample meet this criterion.   

\begin{figure}	
	\begin{center}
		\includegraphics[width=95mm]{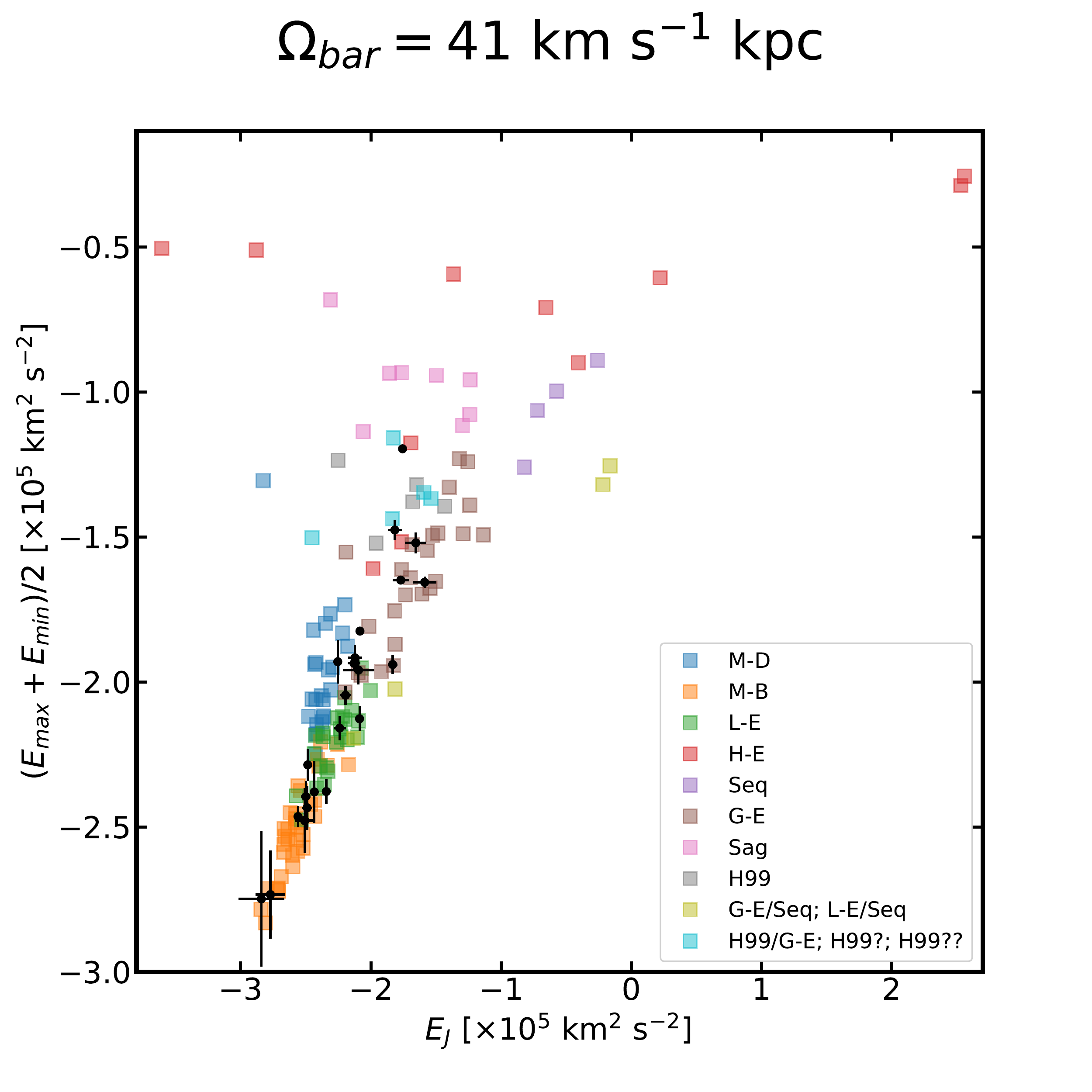}
		\caption{Characteristic orbital energy (($E_{max} + E_{min}$)/2) versus the orbital Jacobi constant ($E_{J}$) in the non-inertial reference frame where the bar is at rest. Square symbols refer to Galactic GCs, colour-coded according to their association with different progenitors from \citep{Massari2019}. The black dots with error bars refer to the Al-rich star analysed in this study.}
		\label{Figure6}
	\end{center}
\end{figure}	

\section{Comparisons to \texttt{ASPCAP}}			

\begin{figure}	
	\begin{center}
		\includegraphics[width=95mm]{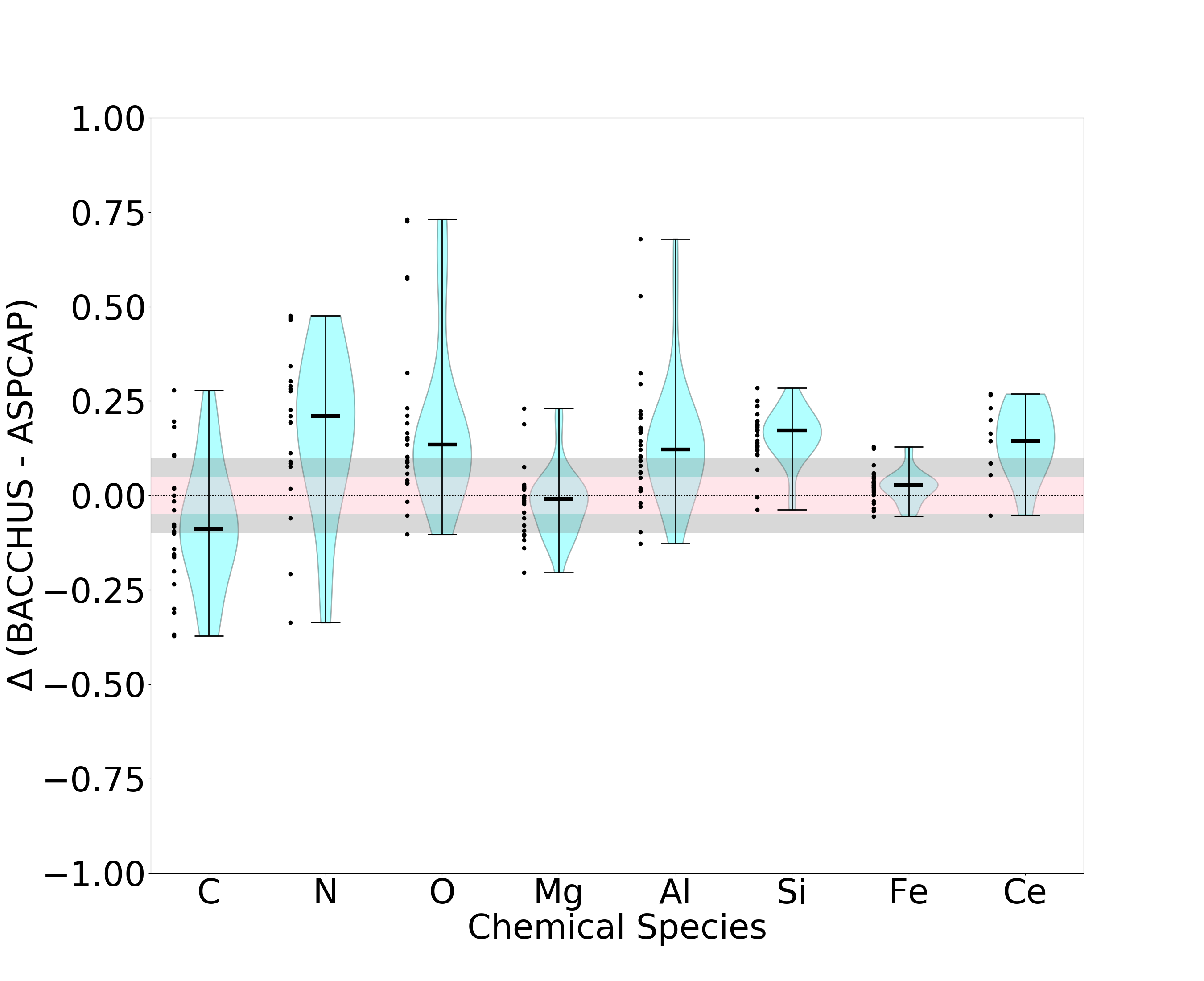}
		\caption{Violin diagram showing the differences between the derived elemental abundances between our analysis with the \texttt{BACCHUS} code against those values of the \texttt{ASPCAP} pipeline for our 29 Al-rich stars. Each violin representation indicates with horizontal lines the median and limits of the distribution. The horizontal shaded regions show the typical errors of $+0.05$ dex, and $+0.1$ dex found by the \texttt{ASPCAP} pipeline.}
		\label{Figure5}
	\end{center}
\end{figure}

Figure \ref{Figure5} we compares our abundance determinations with those of APOGEE DR16 \citep{Ahumada2020}. The differences are shown as a density estimation in a violin representation, whereby the median, min, and max of the differences of each chemical species are marked. We find that the [Fe/H] metallicities we derive are in excellent agreement with \textit{ASPCAP} pipeline, with a median distribution slightly higher than the reported by DR16, but within the typical error ($\pm 0.05$ dex) determined by the \texttt{ASPCAP} pipeline. Furthermore, this indicates that we reproduce the precision of the \texttt{ASPCAP} pipeline for the iron lines. However, there are significant differences in the range of certain abundance ratios comparing \texttt{BACCHUS} to \texttt{ASPCAP}. For example, [N, O, Al, Si, Ce/Fe] ratios derived with \texttt{BACCHUS} are $\sim$0.1--0.75 dex higher than the \texttt{ASPCAP} values, while [O, Mg/Fe] are $\sim$0.05--0.4 dex lower than \texttt{ASPCAP} determinations, possibly due to some issues with the accuracy (zero-point) of \texttt{ASPCAP} abundances, limits of the model grid \citep{Henrik2018}, and/or the difficulty of fitting lines where the intensity is comparable to the variance. For a more detailed discussion, we refer the reader to \citet{Masseron2019}. While the \texttt{ASPCAP} pipeline uses a global fit to the continuum in the three detector chips independently, the \texttt{BACCHUS} pipeline places the pseudo-continuum in a region around the lines of interest. We believe that our manual method is more reliable, since it avoids possible shifts in the continuum location due to imperfections in the spectral subtraction along the full spectral range. 

\begin{figure}	
	\begin{center}
		\includegraphics[width=95mm]{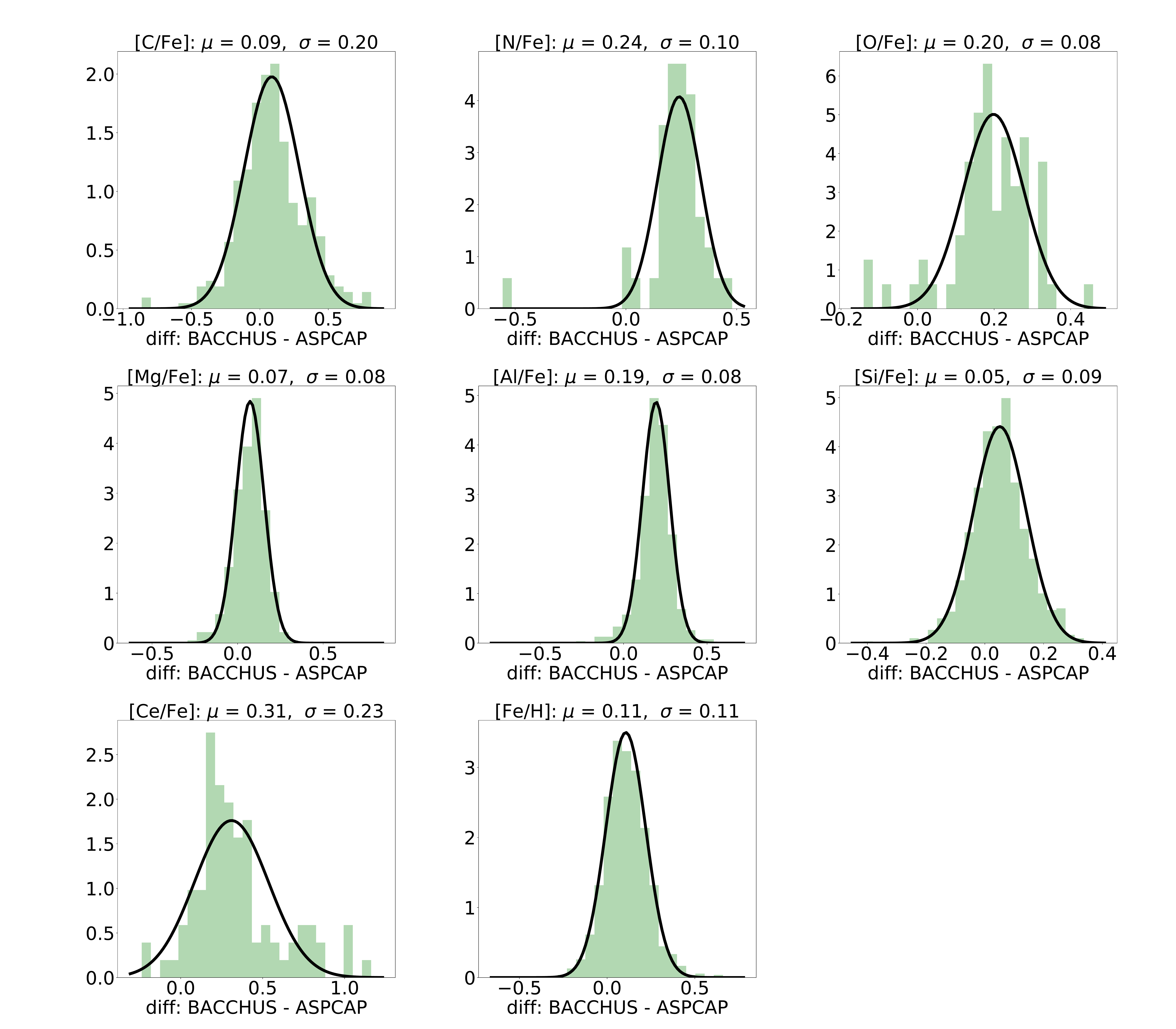}
		\caption{Differences in abundances produced by the \texttt{BACCHUS} and \texttt{ASPCAP} for a control sample of $\sim1,000$ giants belonging to the main body of the MW (halo, disk and bulge), at the same metallicity range as examined in this study. The numbers in each title of each panel indicate the peak ($\mu$) and the standard deviation ($\sigma$) of the fitted normal distribution, showing that \texttt{ASPCAP} values are significantly offset from the \texttt{BACCHUS} ones.}
		\label{Figure7}
	\end{center}
\end{figure}

Figure \ref{Figure7} shows the typical offset of each chemical species between the \texttt{BACCHUS} and \texttt{ASPCAP} pipeline for a control sample of $\sim$1,000 metal-poor stars ($-2\lesssim$[Fe/H]$\sim-0.7$) belonging to the main components of the MW (halo, disk, and bulge). We find that \texttt{ASPCAP} 
significantly underestimates most of the chemical species by about $\sim$0.1 to 0.3 dex for most of the metal-poor stars \citep[although see][for an alternative view]{Nataf2019}. Such offsets were taken into consideration for the whole MW stars presented in Figure \ref{Figure1}.

\section{Al I features}	

\begin{figure*}	
	\begin{center}
		\includegraphics[width=190mm]{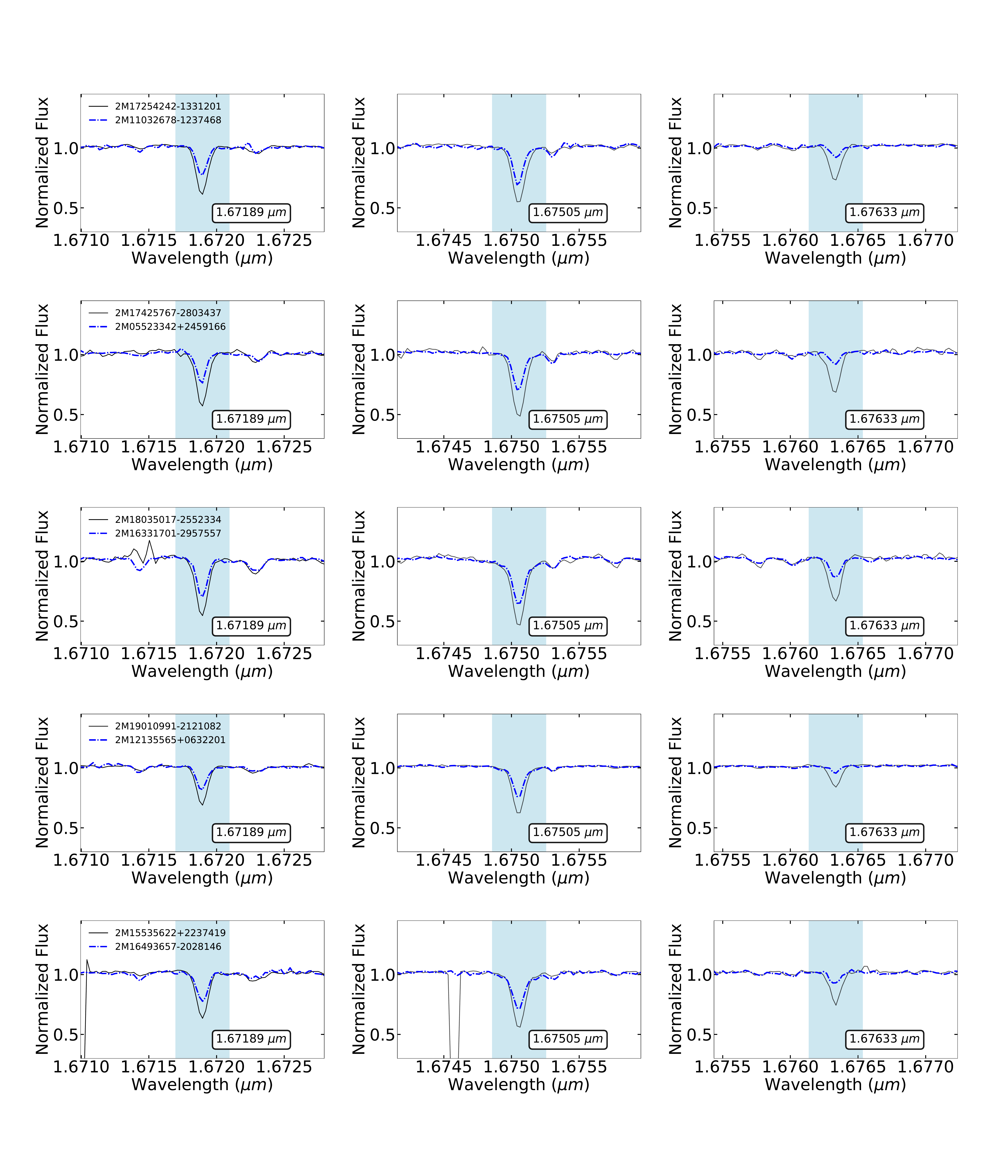}
		\caption{Comparison between the spectra of a normal (blue) and an Al-rich (black) star, with similar stellar parameters around Al I absorption lines at 1.67189, 1.67505 and 1.67633 $\mu$m. The last row shows the typical spectrum of a bulge NRS (blue) with low aluminum enrichment from \citet{Schiavon2017}.}
		\label{Figure4}
	\end{center}
\end{figure*}

Figure \ref{Figure4} provides a brief examination of typical \textit{H}-band spectra for five Al-rich stars and Al-normal stars that exhibit clear Al I spectral absorption features. This figure reassures us of the existence of a real chemical peculiarity in the rich stars. The spectra of these stars are compared in a wavelength range containing several Al I lines, i.e., at 1.67189, 1.67505 and 1.67633 $\mu$m, which are indicated by the cyan shadow region. The Al-rich stars have remarkably stronger Al I lines, compared to Al-normal stars with the same atmospheric parameters. This can only be attributed to a difference in aluminum abundance.

\end{appendix}

\end{document}